\documentstyle[11pt]{article}

\begin{document}
\begin{flushright}
astro-ph/9401048
\end{flushright}
\begin{center}

{\bf Scale-Invariant Spectrum of Cosmic Background Radiation as a
     Feature of the Universe with Negative Curvature}
\vspace{0.2in}

V.G.Gurzadyan$^{\dag\ddag}$
\footnote{E-mail address: gurzadyan@vxrmg9.icra.it}
and
A.A.Kocharyan$^{\#}$
\footnote{E-mail address: armen@vaxc.cc.monash.edu.au}
\vspace{0.2in}

$^{\dag}$ International Center for Relativistic Astrophysics,
Dipartimento di Fisica,
Universita di Roma ``La Sapienza", 00185, Rome, Italy
\vspace{0.1in}

$^{\ddag}$ Department of Theoretical Physics, Yerevan Physics Institute,
375036 Yerevan, Armenia
\vspace{0.1in}

$^{\#}$ Department of Mathematics, Monash University ,
Clayton, Victoria 3168, Australia
\end{center}

{\bf  As was shown before (Gurzadyan and Kocharyan, 1992, 1993ab)
the statistical properties (exponential mixing) of motion of CMB
photon beams in Friedmann Universe with negative curvature  can
have definite observable  consequences  including the decrease
of the CMB anisotropy amplitude after the last scattering,
distortion of images on CMB sky maps.
Here we consider one more consequence concerning the behaviour
of temperature autocorrelation function $C(\theta, \beta)$,
where $\theta$ is the modulation (separation) angle,
and $\beta$   the observing beam effective angle.
We prove  that in the Universe with negative curvature any
spectrum of perturbations at last scattering epoch becomes a
scale-invariant by present time.
It is  also predicted that the temperature autocorrelation function should
increase while obtained from measurements with small beam sizes within
some {\it mixing beam angle}.}
\vspace{0.1in}


\newpage

The discovery of anisotropy of the Cosmic Microwave Background radiation
by Cosmic Background Explorer (COBE) announced the beginning of far more
informative studies on the evolution and the present structure of the
Universe.
The first and most popular viewpoint of COBE and other available data
concerning in particular the temperature autocorrelation function
$C(\theta,\beta)$ was the support of Harrison-Zeldovich scale-invariant
spectrum of initial fluctuations and hence of inflationary scenario
(Smoot et al., 1992; Hancock et al., 1994).

Below we consider another possible interpretation of the same data,
namely as indication of the negative curvature of the Universe.
Our approach is based on the studies on properties of
photon beam motion in Friedmann-Robertson-Walker Universe with $k=-1$,
which can be reduced to the problem  of the
behaviour of correlation function of geodesic flow on
3-manifold with negative constant curvature. In particular the
study of the exponential decay of
correlation functions in the context of  CMB problem has led to the
prediction of the following observable effects
(Gurzadyan and Kocharyan, 1992, 1993ab):
a) decrease of CMB anisotropy after the last scattering epoch
depending on the value of the density parameter $\Omega$;
b) distortion of images of anisotropy structures on CMB maps;
c) possibility of measuring of similar properties of CMB photon beams
arriving from regions having no causal connection.

The first effect as follows from the calculations
(Gurzadyan and Kocharyan, 1992, 1993a),
can lead to a damping of
$\delta T/T$ since last scattering epoch, $z=1000$, up to the present time
by a rather significant factor depending on $\Omega$
(up to $100-1000$ at $\Omega\sim0.1-0.3$).

The second effect can lead to complex-elongated structures of the constant
temperature curves on CMB maps.
As a simplified indicator, the degree of elongation, if measured on
CMB maps can give the present value of $\Omega$ by the formula derived in
(Gurzadyan and Kocharyan, 1993b).
The preliminary statistical analysis of COBE-DMR maps (Gurzadyan and Torres,
1993)
does reveal existence of elongated shapes; if attributed to the described
effect the results indicate $\Omega=0.2-0.4$; special
pattern-recognition code was developed for that study (Torres, 1993)
to distinguish from other map properties (Martinez-Gonzalez and Cayon, 1993).
The third effect, equivalent to the fact that {\it two ink drops are mixing
in a cup of water in the same manner without requiring information on each
other}, removes the mystery of the horizon problem.

All these three effects occur
only in Friedmann Universe with negative curvature (i.e. they
disappear if either $k=0$ or $k=+1$), and therefore can be of particular
importance for probing of this principal property of the Universe.

Evidently nothing is happening to any individual photon during free
propagation after the last scattering, and these effects are purely
statistical and are determined by the principal limitations of
obtaining information during measurements, i.e. by the  impossibility of
reconstruction of the trajectory of given
photon while observing within finite smoothing angle due to the overlapping
of exponentially deviating geodesics in any cut of phase space.
This situation of
loss of information can be described via {\it random sequences} of
Kolmogorov (1969) and Chaitin (1992).

Below we consider another aspect of the problem of exponential decay of
correlators of geodesic flow having particular observational
consequences concerning the properties of temperature  autocorrelation
function  $C(\theta, \beta)$ (Silk, 1992).

Our present study  is essentially determined by recent rigorous
results on the correlation functions of geodesic flow on 3-manifold with
negative curvature (Pollicott, 1992).
Geodesic flow on $d$-dimensional manifolds with negative curvature
has been object of intense
studies rather long time, since Hadamard (1898), Hedlund, Hopf (30ies) up to
fundamental results by Anosov  and others.
In particular,  it was proved that  geodesic flow on
closed (compact, without boundary) manifold with negative curvature
is Anosov system (Anosov, 1967),
is possessing the strongest chaotic properties (e.g. mixing of all degrees),
with positive Kolmogorov-Sinai (KS) entropy.
Rather informative characteristics of any dynamical system are the
correlation functions, in particular the law of their decay, though the
obtaining of rigorous results for each given system usually requires
great efforts and in fact for only a few classes of dynamical systems
results on correlation functions exist up to date. The new
result (Pollicott, 1992) being of remarkable interest for us reads:
the correlation function
$$
   b_{A_1,A_2}({\lambda})=\int_{SM}A_1\circ f^{\lambda}\cdot A_2 d\mu
         -\int_{SM}A_1d\mu\int_{SM}A_2d\mu
$$
of the geodesic flow $\{f^\lambda\}$ on the unit tangent bundle $SM$ of the
Riemann 3-manifold $M$ with negative constant curvature
decreases by exponential law, namely for all functions $A_1,A_2\in L^2(SM)$
but a finite-dimensional space in $L^2(SM)$ there exists $c>0$ such that
\begin{equation}
   \left|b_{A_1,A_2}(\lambda)\right|
      \leq c\cdot \left|b_{A_1,A_2}(0)\right|\cdot e^{-h\lambda} \ ,
\label{expmix}
\end{equation}
where $\mu$ is the Liouville measure and $\mu(SM)=1$,
$h$ is the KS-entropy of the geodesic flow $\{f^{\lambda}\}$.

We now take up the question of how one describes
the free motion of photons
in the $4D$-World (for details see (Gurzadyan and Kocharyan, 1993a)).
To make matters more
simple we will only
consider a Friedmann $4D$-World $W$, decomposed into
a homogeneous $3D$-Universe characterized by some coordinates $x$,
evolving with respect to the cosmic time $t$.
Thus each event in the World can be assigned unique
space-time coordinates $(x,t)$. One can then project any trajectory
$\gamma$ from the World $W$ into the Universe $U$ simply by associating to
$\gamma(\lambda)=(x(\lambda),t(\lambda))$ the curve $c(\lambda)=x(\lambda)$.
It turns out that null geodesics in the World project onto geodesics in
the Universe with new affine parameter:
$$
  \lambda(t)=\int_{t_0}^t \frac{ds}{a(s)} \ .
$$
The latter relates to the problem of ``internal time" of
the $K$-systems in cosmology considered initially in (Lockhart et al., 1982).
As it can be easily shown by means of the formulae
in (Peebles, 1993; Gurzadyan and Kocharyan, 1993a), the degree of smoothing
of anisotropy $\delta T/T$ in post-scattering epoch is determined by
the KS-entropy $h$ and affine parameter $\lambda$ which depend on the
parameters of the Universe. For the matter-dominated Universe one has
\begin{equation}
   e^{h\lambda}=
      (1+z)^2\left[\frac{1+\sqrt{1-\Omega}}{\sqrt{1+z\Omega}
                          +\sqrt{1-\Omega}}\right]^4 \ .
\label{factor}
\end{equation}
For $z\gg\Omega^{-1}$ one may obtain
\begin{equation}
   e^{h\lambda}\sim\frac{(1+\sqrt{1-\Omega})^4}{\Omega^2} \ .
\label{factorApr}
\end{equation}
Here our aim is to investigate the behaviour of
the temperature autocorrelation function $C(\theta,\beta)$
using properties of the correlation function of the geodesic flow
$\{f^{\lambda}\}$.

One may readily see that if
$$
  A_1(u)={\cal T}(u)-1=\frac{T(u)}{\bar{T}}-1\ ,
$$
and
$$
  A_2(u)=\frac{\chi_{{\cal K}(v)}(u)}{\mu({\cal K}(v))}-1 \ ,
$$
where $T(u)$ is the real temperature at decoupling time at $u\in SM$,
$$
  \bar{T}=\int_{SM}Td\mu\ ,
$$
${\cal K}(v)\subset SM$ is a  beam of detected photons at $v\in SM$,
and  $\chi_{{\cal K}(v)}$ is the characteristic function
of the set ${\cal K}(v)$,
then normalized ``measured" temperature at point $u$ at $\lambda$
defines as follows
\begin{equation}
  {\cal T}_{\lambda}(u)=\frac{1}{\mu({\cal K}(u))}
                         \int_{{\cal K}(u)}{\cal T}\circ f^{\lambda}d\mu
                       = 1+b_{A_1,A_2}({\lambda}) \ .
\label{meanT}
\end{equation}
Therefore one may using Eqs.~(\ref{expmix}), (\ref{factor}) arrive at
the following inequality
$$
  \left|{\cal T}_{\lambda}(u)-1\right|
         \leq c \cdot \left|{\cal T}_0(u)-1\right|
         \cdot \frac{1}{(1+z)^2}
          \cdot\left[\frac{\sqrt{1+z\Omega}
                          +\sqrt{1-\Omega}}{1+\sqrt{1-\Omega}}\right]^4\ .
$$
Thus we see, that for any $u$, i.e. at any space point in any sky direction,
{\it the ``measured" temperature tends to the constant
mean temperature, i.e. the isotropic state is the final state}.

For normalized temperature autocorrelation function
\begin{equation}
 {\cal C}_{\lambda}(\theta,\beta)
     =\langle {\cal T}_{\lambda}(u)
              {\cal T}_{\lambda}(v) \rangle_{g(u,v)=cos\theta}\ ,
\label{C}
\end{equation}
the following inequality holds
$$
  \left|{\cal C}_{\lambda}(\theta,\beta)-1\right|
  \leq  c \cdot \left|{\cal C}_0(\theta,\beta)-1\right|
          \cdot\frac{1}{(1+z)^2}
          \cdot\left[\frac{\sqrt{1+z\Omega}
                          +\sqrt{1-\Omega}}{1+\sqrt{1-\Omega}}\right]^4 \ .
$$
The following conclusion immediately follows from this formula :

{\it Temperature autocorrelation function $C(\theta,\beta)$
is almost constant with respect to $\theta$ at the present
time regardless on the initial perturbations at $z\sim1000$.}

Numerically the effect determined by Eq.~(\ref{factor}) as was
mentioned above is large enough for CMB
photons (Gurzadyan and Kocharyan, 1993a),
to reveal it during the evolution of the Universe due to
negativity of curvature of the Universe.

Compare this conclusion with  COBE-DMR and other available data.
The scale invariant spectrum is
considered as the essential feature of those data and is interpreted as
confirmation of Harrison-Zeldovich spectrum of initial perturbations
at the last scattering epoch.
Here we proved that in Friedmann Universe with $k=-1$ any spectrum
should become nearly scale invariant since last scattering surface.

Using Eq.~(\ref{meanT}) and the fact that the geodesic
flow is an Anosov system
one may show that there exists an angle $\phi$ such that
smoothing factor $s$ of $\delta T/T(\beta)$ is
almost constant
if either $\beta\gg\phi$ ($s\sim e^{-h\lambda}$)
or $\beta\ll\phi$ ($s\sim 1$),
and increases as
$\beta$ is decreased at  $\beta\sim\phi$ by the law
$s\sim const/\beta\cdot e^{-h\lambda}$.

One can explain this result and the content of the {\it
smoothing angle} $\phi $ as follows.
As we have mentioned above,  the smoothing
of CMB is a statistical effect due to the loss of information during the
measurements while averaging within some beam size. Therefore, the more narrow
is the beam size (pencil-beam), the less information is lost at smoothing
in terms of the temperature autocorrelation function. Maximal
information, e.g. the anisotropy corresponding exactly the last
scattering surface should be obtained while measuring by beams within
some  mixing angle $\phi$; measurements on small
angles should not influence the autocorrelation function which should be
determined solely by physical conditions of the last scattering surface
(its thickness, the Silk effect, etc).

The results  of  at least two experiments with smaller beam size than COBE,
the MAX (Gundersen et al., 1993),
$\delta T/T\leq 4.2\cdot 10^{-5}$, with beam size
$0.5^{\circ}$ and ARGO (De Bernardis et al., 1993),
$ 1.6\cdot 10^{-5}<\delta T/T<2.6\cdot 10^{-5}$, beam size,
$0.8^{\circ} (52')$,
show reliable increase of  anisotropy
compared with  COBE and therefore seem to support our prediction;
somewhat less supporting is the result of
South Pole ACME experiment (Gaier et al., 1992),
$\delta T/T\leq 1.4\cdot 10^{-5}$,
with  $1.5^{\circ} $ beam size.  Evidently
more accuracy  and statistics is needed to distinguish
our prediction from, say Doppler shift and predictions of
certain CDM galaxy formation models.

This prediction, therefore
indicates a way to obtain the sign of the curvature of the Universe,
the real anisotropy  at the last scattering epoch as well as the present
value of $\Omega $.

To summarize,  the effects we are considering are model-independent,
i.e. they do not depend on dark matter models, initial fluctuations, etc.
The effects should appear if the  Universe is Friedmann
- homogeneous and isotropic, and has negative curvature.
One has a  real possibility of getting
information directly on the curvature of the Universe, since the effects
disappear at $k=0$ and $k=+1$.
\vspace{0.1in}

{\bf  Acknowledgements}

We thank I.Prigogine, A.Polnarev and D.W.Sciama for
stimulating discussions, P. de Bernardis, F.Melchiorri and S.Torres
for explanation of observational details. One of us (V.G.G.) is thankful
to R.Ruffini for hospitality in ICRA.

\end{document}